# The ARGUS Vertex Trigger

N. Koch, M. Kolander, H. Kolanoski[*], T. Siegmund

Institut für Physik[†], Universität Dortmund, Germany

J. Bergter, P. Eckstein, K.R. Schubert, R. Waldi

Institut für Kern- und Teilchenphysik[‡], Technische Universität Dresden, Germany

M. Imhof, D. Reßing[§], U. Weiß, S. Weseler

Institut für Experimentelle Kernphysik[¶], Universität Karlsruhe, Germany

## Abstract

A fast second level trigger has been developed for the ARGUS experiment which recognizes tracks originating from the interaction region. The processor compares the hits in the ARGUS Micro Vertex Drift Chamber to 245760 masks stored in random access memories. The masks which are fully defined in three dimensions are able to reject tracks originating in the wall of the narrow beampipe of 10.5 mm radius.

## 1 Introduction

In 1991 a new vertex detector was installed into the ARGUS experiment [1] which should allow tagging of heavy flavour decays by a precision reconstruction of secondary vertices. The system consists of a high resolution drift chamber [2] and a silicon strip detector [3]. Together with the installation of these components the radius of the beam pipe around the interaction region was decreased to only 10.5 mm, in order to keep the lever arm of the reconstructed tracks to the vertex as short as possible. Aperture limiting elements within the beampipe move even as close as 9 mm to the beam axis.

However, as a result of the narrow beampipe the background rate from particles scattered in the wall of the beam pipe near the interaction region increased dramatically. With the previously used trigger settings the trigger rate of the experiment went up by roughly an order of magnitude while the previous rate of $\sim 10\,\mathrm{Hz}$ already causes a deadtime of about 10%. Therefore, in order to maintain a high efficiency for annihilation processes it became necessary to significantly reduce the background trigger rate. Since most of the background tracks originate relatively close to the interaction region an efficient rejection demands that the track origin has to be sufficiently well reconstructed on the trigger level.

[*]now at Institut für Physik, Humboldt Universität, Berlin, Germany

[†]Supported by the German Bundesministerium für Forschung und Technologie, under contract number 054DO51P.

[‡]Supported by the German Bundesministerium für Forschung und Technologie, under contract number 056DD11P.

[§]now at DESY, Hamburg, Germany

[¶]Supported by the German Bundesministerium für Forschung und Technologie, under contract number 055KA11P.

Figure 1 shows a cross section of the beam pipe in the ARGUS interaction region. On top of the figure the distribution of measured event vertices along the beam line ($z$ direction) is displayed. Only the peak around $z = 0$ is due to the wanted events. Prominent peaks show up at the positions of the scrapers shielding the synchrotron radiation at the entrance to the narrow beam pipe part. Another peak is due to a shielding of the silicon detector electronics (which was only installed part-time). In a perpendicular cross section ($r - \phi$ view, not shown) one observes the radial distribution of the event vertices enhanced at the radius of the beam pipe. A quantitative analysis of the vertex distributions suggests that an efficient background rejection can be achieved with a vertex trigger which resolves the impact of a track to the beam line ($d_0$) better than 10 mm and the $z$ projection within about 50 mm.

The ARGUS Micro Vertex Drift Chamber ($\mu$VDC) was designed to allow for a full three-dimensional track reconstruction, with similar emphasis on both the $r - \phi$ and $r - z$ projections. This chamber can provide all necessary information to allow for a full three-dimensional vertex reconstruction on the trigger level. With a cell size of about 5 mm the resolution achievable using only the hit information of the 1070 wires is sufficient. With the expected rate reduction and the given read-out time a decision time of about 30 $\mu$s can be tolerated.

This paper is organized as follows: in section 2 the detector parts used for the trigger are explained. Section 3 describes the trigger algorithm, section 4 the hardware of the trigger components, and section 5 the hardware and software necessary to control the Vertex Trigger. In section 6 first experimental results on the trigger performance are presented, and the paper ends with a summary in section 7.

## 2 The ARGUS Detector

### 2.1 Overview of the Detector

The ARGUS detector at the $e^+e^-$ storage ring DORIS in Hamburg is optimized for measuring $e^+e^-$ annihilation in the energy range of the $\Upsilon$ resonances.

The inner tracking detectors, a vertex detector system and the main cylindrical drift chamber, are operated in a solenoidal magnetic field of 0.8 Tesla. The component closest to the interaction region is the silicon strip detector at a minimal distance of 1.5 mm to the beampipe which has a radius of only 10.5 mm. Further out follows the Micro Vertex Drift Chamber ($\mu$VDC) which will be described in more detail below since the Vertex Trigger (VT) decision is solely based on the signals from this detector. The main cylindrical drift chamber [4] covers a radial range between 15 cm and 86 cm and is 2 m long. It contains 5940 drift cells (18 × 18.8 mm$^2$) in 36 layers, half of which are stereo layers with angles between 2.3° and 4.6°. The tracking detectors are surrounded by time-of-flight scintillation counters, followed by a lead-scintillator electromagnetic calorimeter, the magnet coil and a muon detector.

In the following we use cylinder coordinates ($r, \phi, z$) with the origin in the nominal interaction point (centre of the detector), where $r$ is the radial distance to the beam line,

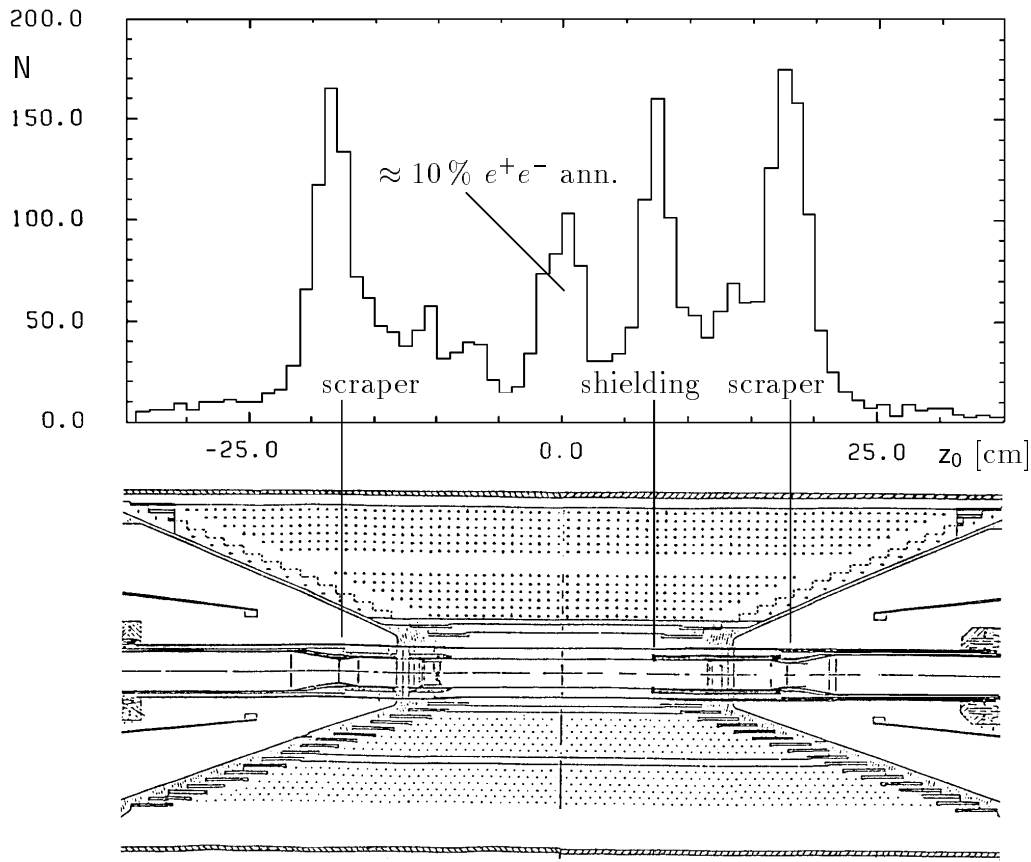

**Figure 1** Top: $z$-distribution of the main vertex of events accepted by the standard track triggers in 1991 after installation of the new vertex detectors into the ARGUS experiment. Bottom: cross section through the new vertex detectors and the narrow part of the beam pipe.

$\phi$ the azimuth around the beam and $z$ the axial coordinate along the beam.

## 2.2 The Micro Vertex Drift Chamber

The design goal of the $\mu$VDC [2] was to allow a reconstruction of secondary vertices with similar resolution in all three dimensions. This was possible by introducing extreme stereo angles of the wires. With a total of 16 layers the chamber consists of four $0°$ layers (1,2,9,10; counted from the innermost layer), six $+45°$ layers (3,5,7,11,13,15) and six $-45°$ layers (4,6,8,12,14,16). The $45°$ wires are strung across five thin beryllium vanes leading to a pentagonal cross section of the chamber (fig. 2). The number of drift cells per layer increases from 35 in the 1st layer to 100 in the 16th layer. In total there are 1070 nearly quadratic cells (5.320 mm × 5.178 mm). The minimal radius $r_{min}$ of each layer varies between about 25 mm and 103 mm and the axial length of each layer between about 150 mm and 600 mm (see fig. 1) matching the acceptance of 93% of $4\pi$. The vanes introduce an inefficiency of the azimuthal acceptance of about 2%.

The wire arrangement and the symmetry of the chamber geometry determine the algorithm developed for the Vertex Trigger. Azimuthally there are five equivalent sectors between the vanes. In each sector the 16 wire layers are parallel planes at equal distance between each other. Since the sensitive volume of the chamber has a pointing geometry

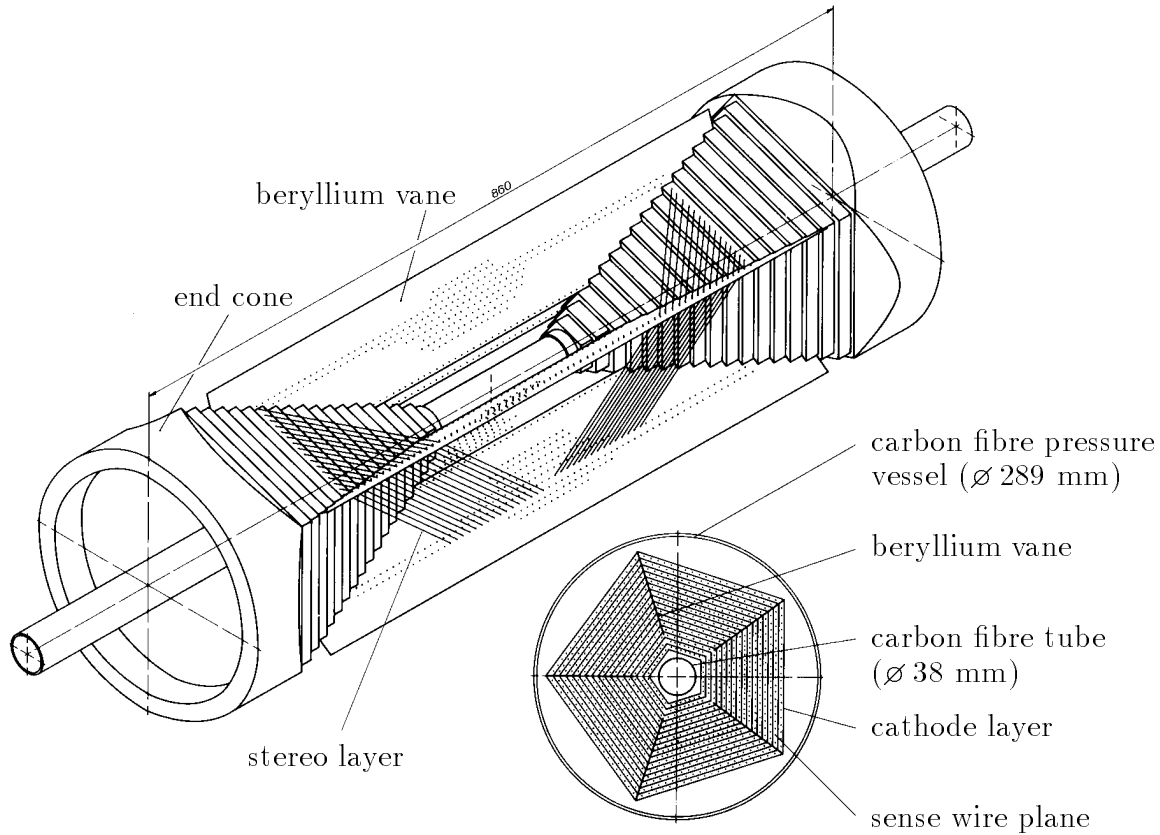

**Figure 2** Schematic view and cross section of the micro vertex chamber.

with respect to the origin the length and width of each layer scale with $r_{min}$.

## 2.3 ARGUS Trigger System

The ARGUS detector is run with a two-level trigger system. The first level trigger (pretrigger) decides within about 300 ns using the fast signals from the time-of-flight counters and the calorimeter. Since the bunch crossing frequency is 1 MHz this trigger generates no deadtime. The pretrigger consists of several subtriggers which use different energy thresholds and different topological requirements: A total energy trigger (ETOT) requiring a large and balanced energy deposit in the calorimeter, a high energy shower trigger (HESH) with a large local energy deposit, and two two-prong triggers requiring either an opening angle in $\phi$ (CMAT), or one track in each $z$ hemisphere (CPPT). For monitoring purposes also a random pretrigger is installed.

Any pretrigger starts the second level trigger processor, the Little Track Finder (LTF) [5], to search for track candidates in the main drift chamber. An ETOT or random pretrigger, however, masks the reject signal of the LTF and these events are recorded independent of the second trigger stage. For monitoring purposes, every $100^{th}$ pretrigger will be accepted via a VETO overriding the second level decisions.

The LTF uses hits within the axial and innermost stereo layers of the main drift chamber to recognize in the $r - \phi$ plane tracks originating from the interaction region.

The track finding algorithm compares the measured hits in the drift cells to masks which are stored in random access memories (RAM). Starting from a cell in a reference layer cells in the other layers are associated to a mask, which covers tracks within a certain transverse momentum ($p_T$) and azimuth angle ($\phi$) range. The second innermost layer of the drift chamber was used as reference layer before the installation of the Vertex Trigger. The processing time of the LTF is typically $20\,\mu$s.

The trigger is fully efficient for tracks with $p_T > 130\,MeV$ and an impact distance to the beam $d_0 < 30$ mm. This impact parameter resolution is not sufficient to discriminate against tracks originating in the new beam pipe wall. In addition, this trigger setup has no possibility to incorporate $z$-information.

# 3 The Algorithm of the Vertex Trigger

## 3.1 General Concept

The Vertex Trigger searches for tracks originating from the interaction region and passing the $\mu$VDC. On the basis of the found tracks the Trigger Processor defines a virtual reference layer to be used by the LTF, or rejects the event if no track has been found. Thus only those tracks can be found by the LTF which have a vertex in the volume resolved by the Vertex Trigger. This scheme implies that the VT and the LTF have to work consecutively. The decision times of both add to about $50\,\mu$s.

The VT algorithm tries to match the hits in the 1070 cells of the $\mu$VDC with patterns of wires which form a mask. A track can be defined by the polar angle $\theta$, the azimuthal angle $\phi$, the distance of closest approach $d_0$, the $z$ coordinate at the point of closest approach $z_0$ and the transverse momentum $p_T$. Each mask defines a range in the track parameters $\theta, \phi, d_0, z_0$. Since the radial extension of the $\mu$VDC is small the track curvature is not taken into account. A track with $p_T = 100$ MeV deviates in the $\mu$VDC from a straight line coming from the origin by less than 6 mm (sagitta). For the definition of the masks in the following we consider only straight line tracks.

## 3.2 Definition of the Masks

Because of the five-fold symmetry of the $\mu$VDC a set of masks is initially defined for one of the five sectors. The masks for the other sectors are obtained by simple transformations of wire numbers.

As explained in section 2.2 the size of each layer in any sector of the $\mu$VDC scales with $r_{min}$. Stretching each layer in both dimensions by a factor proportional to $1/r_{min}$ transforms all layers to the same size. We now define a Cartesian coordinate system $(r, s, t)$ where $r$ remains the minimal distance to the beam ($r = r_{min}$ for each layer), $t$ is the transformed $z$ coordinate with the origin at $z = 0$ and $s$ is the coordinate perpendicular to $r$ and $t$ starting with $s = 0$ at a vane and increasing with the azimuth $\phi$ (fig. 3).

In this coordinate system all straight tracks originating from $r = 0$, $z = 0$ are parallel, vertical lines. A mask selecting such tracks corresponds to a vertical column whose cross section is adjusted to the three wire orientations. This leads to the hexagons shown in fig. 4 which are initially defined without overlap leaving uncovered areas. Each hexagon can then be increased by an overlap factor chosen to be 1.2, which finally defines the cross section of the column.

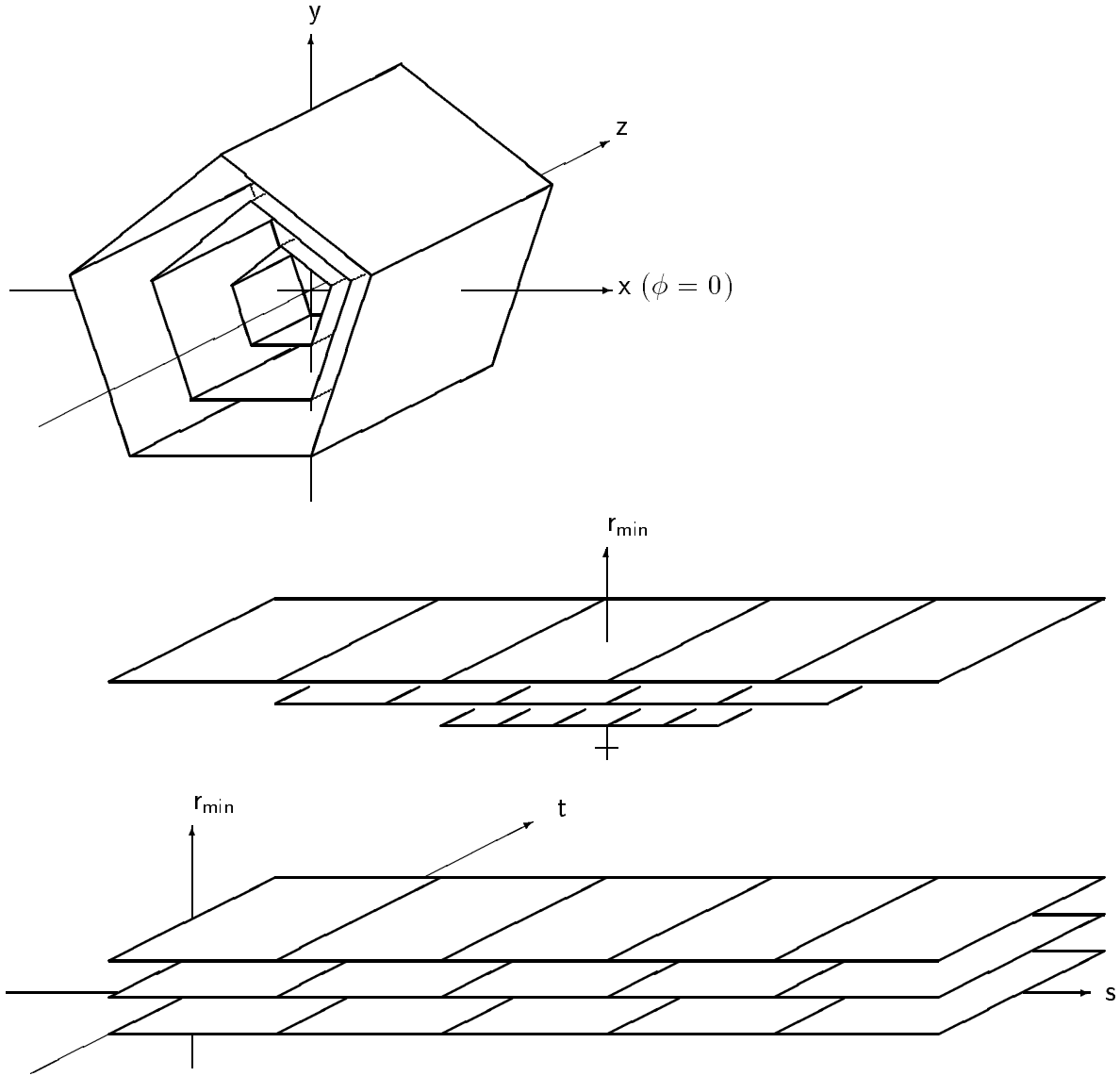

**Figure 3** Definition of the coordinate system. Three layers of the $\mu$VDC are shown to illustrate the transformation from Cartesian coordinates $x, y, z$ to the Vertex Trigger coordinates $r, s, t$.

All drift cells in a plane having a common cross section with a mask form a group. A track candidate corresponds to a pattern of hits with one cell per group in each layer. Since the hexagon widths are smaller than a single physical drift cell, the actual area covered by a mask is larger than the ideal one shown in figure 4, and the whole area—including the white triangles in the figure—is covered completely.

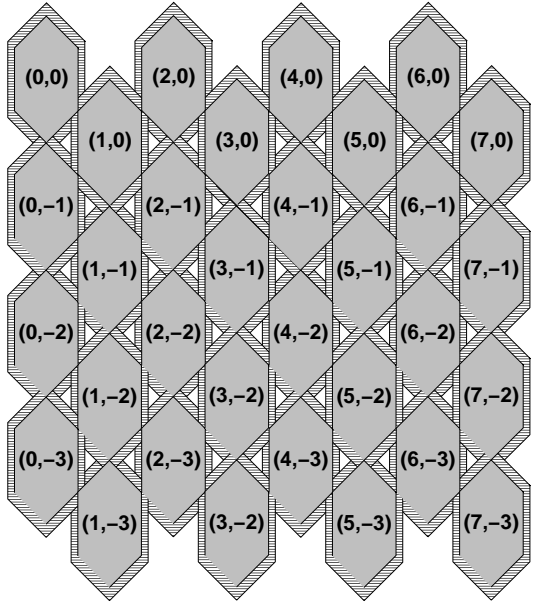

**Figure 4**  Definition of the cross section of the masks in the $s, t$ plane.

A straight track from the interaction point is a line passing all $s, t$-planes perpendicularly, while tracks with different origins deviate in angle, and its intersection with the hexagonal column corresponding to a mask will eventually be too short to be accepted. The width of the masks therefore determines the resolution in the track vertex parameters $z_0$ and $d_0$. Accounting for the length of the interaction region ($\sigma \approx 18$ mm) and maintaining a good vertex resolution requires different sets of masks for different $z_0$ ranges. The definition of the masks can be done as for the $z_0 = 0$ case if each plane is shifted against the neighbouring plane by an amount $\Delta t$ in $t$ direction such that the straight tracks from $z = z_{0i}$ become verticals again.

The mask parameters have been optimized using a simulation program based on preliminary information about the chamber performance, with the aim to maximize the track efficiency and minimize the number of masks [6]. In the actual design phase a solution was found exceeding the minimal requirements from experiment and thereby leaving room for improvements after initial experiences. For the first running with the VT, in $t$ direction only half of each sector was used corresponding to a polar angle range $-0.82 < \cos\theta < +0.82$. This restriction roughly matches the angular acceptance of the LTF. The covered area of each sector is divided into $64 \times 64$ hexagons. The $z_0$ range is $\Delta z_0 = \pm 42$ mm. With 12 $z_0$ mask sets this results in a total of $5 \times 64 \times 64 \times 12 = 245760$ masks for the whole chamber. Each mask corresponds to a set of the four track parameters $\theta, \phi, d_0, z_0$ with an average granularity of $\Delta\phi = 20$ mrad, $\Delta\theta = 40$ mrad, $\Delta d_0 = 7$ mm and $\Delta z_0 = 7.5$ mm.

## 3.3  Organisation of Trigger Decisions

To allow for inefficiencies in the $\mu$VDC not all 16 groups belonging to a mask have to have fired to accept a track candidate. Depending on the number and position of missing layers, we define a quality number for track candidates in a way to compromise efficiency with background rejection. The track quality is derived from the number of groups contributing in each of the three projections, i.e. the 0° and ±45° layers. Since each projection has half of its layers in the inner and half in the outer part of the $\mu$VDC both halves enter separately into the quality definition. This guarantees a reasonable pointing precision due to the available lever arm, and we avoid to accept background from local clusters of hits. For our first running with the VT table 1 shows the definition of the projection quality and table 2 how the track quality is derived from that. The track quality number can be used in a later stage for more complex trigger decisions, such as track multiplicity requirements.

**Table 1** Definition of the projection qualities: "$\sum_{\text{outer}}$" is the hit count of the subset of layers 8-16, "$\sum_{\text{inner}}$" the count in layers 1-7. Dead layers are predefined as being completely set, and appear as fake layers in the table. All combinations that are not listed have quality 0.

| projection | $\sum_{\text{outer}}$ | $\sum_{\text{inner}}$ | projection quality for # of fake layers | | | | | | |
|---|---|---|---|---|---|---|---|---|---|
| | | | 0 | 1 | 2 | 3 | 4 | 5 | 6 |
| 0° | 2 | 2 | 7 | 7 | 7 | 7 | 7 | | |
| | 1 | 2 | 6 | 6 | 5 | 4 | | | |
| | 2 | 1 | 5 | 4 | 3 | 2 | | | |
| | 1 | 1 | 3 | 2 | 1 | | | | |
| ±45° | 3 | 3 | 7 | 7 | 7 | 7 | 7 | 7 | 7 |
| | 2 | 3 | 6 | 6 | 5 | 4 | 3 | 2 | |
| | 3 | 2 | 5 | 4 | 3 | 2 | 1 | 0 | |
| | 1 | 3 | 4 | 3 | 2 | 1 | 0 | | |
| | 2 | 2 | 3 | 2 | 1 | 0 | 0 | | |
| | 3 | 1 | 2 | 1 | 0 | 0 | 0 | | |
| | 1 | 2 | 1 | 0 | 0 | 0 | | | |

**Table 2** Definition of the track quality number (2 bit) as a function of the projection qualities.

| sum of the 3 projection quality numbers | quality of each projection | track quality |
|---|---|---|
| ≥ 15 | > 0 | 3 |
| ≥ 11 | > 0 | 1 |
| ≤ 10 | – | 0 |

The VT decision depends on the pretrigger type. There are no further requirements if an ETOT, random, or VETO trigger was set. HESH, CPPT and CMAT require at least one track of quality 3, CPPT and CMAT in addition one track of quality 1 with 90° separation in azimuth.

Besides the trigger decision the most important information delivered by the VT is the definition of a virtual reference layer to be used by the LTF (see section 2.3). Since the LTF works in the $r - \phi$ projection on cylindrical layers the hits in the reference layer have to be given by their $\phi$ values. Each $\mu$VDC sector is divided into 16 $\phi$ bins of equal size leading to a total of 80 virtual drift cells in the reference layer. Since in most cases more than one mask is set per track, a central $s$ coordinate is derived from a cluster finding procedure and transformed into a $\phi$ coordinate. A virtual cell in the reference layer is set if a cluster of masks in the corresponding $\phi$ bin is found.

# 4 The Trigger Processing Hardware

Constraints to system design from the experiment are a maximum decision time of at most $30\mu s$, about 100000 masks to achieve the necessary resolution and background rejection in the defined fiducial volume, a given format of the available wire data from the TDC latches and the specified output format to the existing ARGUS trigger.

The reconstruction of a track from projected data with wide stereo angle results in a completely non-local problem, but the symmetries of the chamber suggest a repetitive execution with simple shifts of the input wire data with essentially hard-wired logic. In this spirit the decomposition into smaller subprocesses, which each may be cast into a set of identical electronic modules, is straightforward:

1. Transfer the wire data and store them in a way allowing easy direct access to any hit of each layer.

2. Retrieve chunks of the wire data and map them onto wire groups of suitable granularity which are the elementary building blocks for the masks in one layer and are identical for a whole subset of masks.

3. The processing of these building blocks has to proceed in a physical network representing to a large extent the 3-dimensional geometrical structure of the chamber within a local, connected field of masks. One mask field corresponding to a region in parameter space is processed in parallel.

4. Combine results of a mask field and store intermediate results required for the final evaluation.

5. Compute a trigger decision and data for the LTF from the stored information, and transmit the results.

Steps 2 to 4 may be repeated as often as necessary within the time limit to cover the whole parameter space of masks. The symmetry properties of the chamber keep the complexity of the logic moderate. The actual implementation of this hybrid parallel and serial scheme is constraint by technological, engineering and budgetary considerations. The most important will be discussed below.

The most non-uniform operation is the mapping of wires to wire groups because there are sixteen different layers and various sections of the chamber to be treated and only here single bad wires may be masked. This can be accomplished by casting all logic into look-up-tables in static random access memory (SRAM). The local mapping pattern is addressable by the bit pattern of a small section of a layer and a page number. The different pages contain all patterns used in this local environment. To find the proper page for a given cycle the processing has to be strictly synchronous. Covering a larger region in one layer in parallel can be accomplished by a simultaneous mapping of other wires and concatenating the results by a simple OR operation. The availability of SRAMs with appropriate price, size and speed was one of the most severe constraints for the choice of technology (TTL, CMOS), cycle frequency (about 30 MHz for a single operation) and width of a mask field (both number of wires and groups evaluated in parallel).

Another severe limit is due to step 3. The network is cast into cabling and printed traces. Merely the physical size of standard cables and connectors and the wiring topology constrain the size of the printed circuit boards, the segmentation into modules and again the width of the mask field. In addition we were led to add an intermediate step on the level of projections.

As a compromise we found a very flexible solution that even surpasses the initially defined experimental requirements leaving room for optimizations after first experiences. The track finding has been divided into 480 serial cycles in a pipelining scheme. The complete process of logic operations and data tranfers has been divided into a sequence of small steps that all can be executed in less than 40 ns. These consist mostly of SRAM mappings, simple ORs, data transmissions and delay adjustments. While the data of one mask field are processed in a certain step, the next field is processed simultaneously at the preceding step in a pipeline, thus reducing the overall execution time.

In this setup the number of groups per layer used is 32 resulting in 512 masks processed in parallel. The definition and decomposition of such a mask field is shown in fig. 5. It covers 1/16 of a chamber sector or, with five sectors, 1/80 of the whole chamber (see fig. 5, left). Because of the restriction to central polar angles as explained in section 3.2 only the inner eight fields of a sector are used. Since the mask fields are defined for 12 $z_0$ intervals this leads to the $12 \times 40 = 480$ serial cycles.

The high flexibility stated above is achieved because most of the logic is fully programmable by downloading mapping patterns into the SRAMs. Any change in mask definitions and quality criteria, any temporary change of chamber efficiencies and any change in trigger requirements is easily accomplished, the only hardwired routing being a reflection of the mere geometry of the chamber.

The whole trigger process is now decomposed into subprocesses residing in specialized hardware processors as shown in fig. 6. Each hardware processor is again divided into submodules executing a fraction of the task completely in parallel. All modules are controlled and tightly synchronized by a central instance (main control unit MCU). There are four different hardware processors.

1. The Layer Processor (LP) receives and stores the sorted wire data. During the pipelined processing it maps the hit wires onto the corresponding wire groups. There are 32 groups per layer processed in parallel.

2. The Projection Processor (PP) joins the wire group data of each projection which come from 6 (4) LP modules according to their geometric relation. A data reduction is achieved by mapping the 6 (4) bits onto a 3 bit quality number according to table 1. In addition it provides a fan-out for the next stage to facilitate wiring.

3. The Mask Processor (MP) combines the projection quality numbers of the three projections according to their geometric relation and determines the two bits of track quality according to table 2 for each of the 512 masks represented in hardware. An output stage provides a global OR and a projective OR along $t$ onto $s$.

4. The Trigger Processor (TP) concludes the projective OR along $t$ in slices of $z_0$ by accumulating corresponding cycles. Concurrent pipelined operations provide a cluster search in $s$, a projection onto a virtual reference layer for the LTF, topological track coincidences and global ORs in several levels.

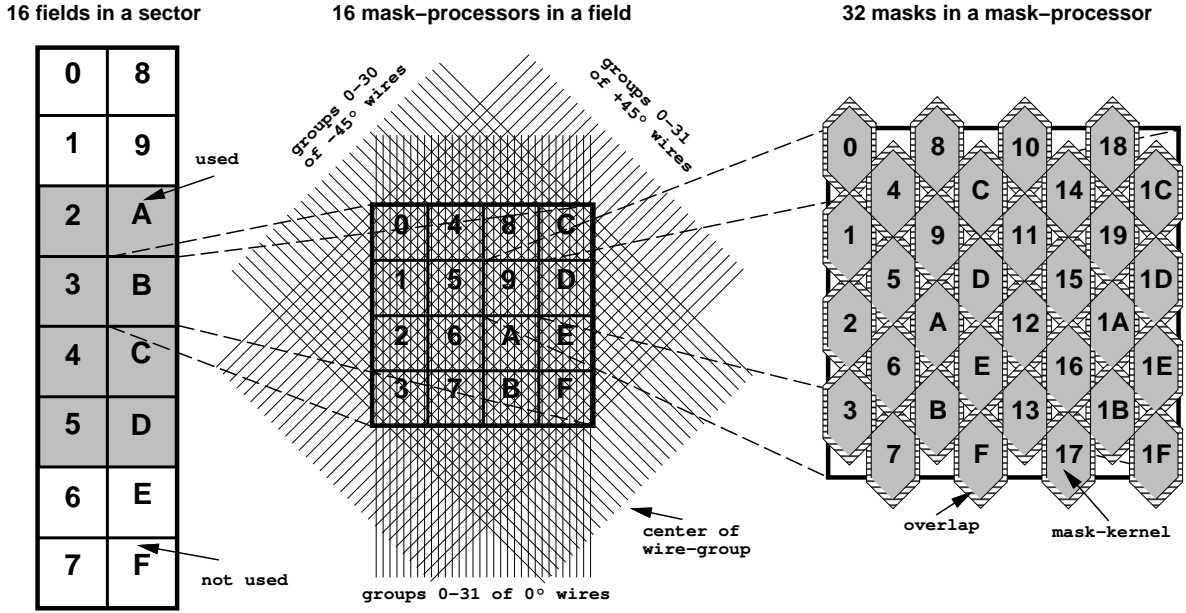

**Figure 5** Left: Definition of mask fields in a chamber sector; centre: dividing a field for the 16 Mask Processors; right: the 32 masks in a sixteenth of a field.

A detailed decomposition of the pipelined process for one mask field is shown in fig. 7 and 8, and a more specific description of the processor hardware is given below.

## 4.1 VME Interface

All processor modules (except the main control unit) have a minimal VME [7] slave interface to allow A24/D32 read and write access to all RAMs and data registers in all access modes of VME specs. Rev. C. In addition to the necessary address registers/counters and data transceivers it consists of only one address decoder and a finite-state-machine in a PAL22V10 to control all internal and external timings and protocols. Due to the nested register and RAM structure of the pipeline the internal timings and the data paths for a VME access are more involved than the pipeline itself. Thus most of the components on board are dedicated to this task.

## 4.2 Description of the Components in the Processor Pipeline

All Trigger Processors consist of several electronic boards and are located in different crates. An overview of the processors in the pipeline is sketched in fig. 6. After a serial loading of the chamber data into the pipeline the data are processed by the four processors defined above. The data flow through all components is shown in fig. 7. Each pipeline fill has another cycle number which is synchronously changed by the pipeline clock. A detailed description is given in [8].

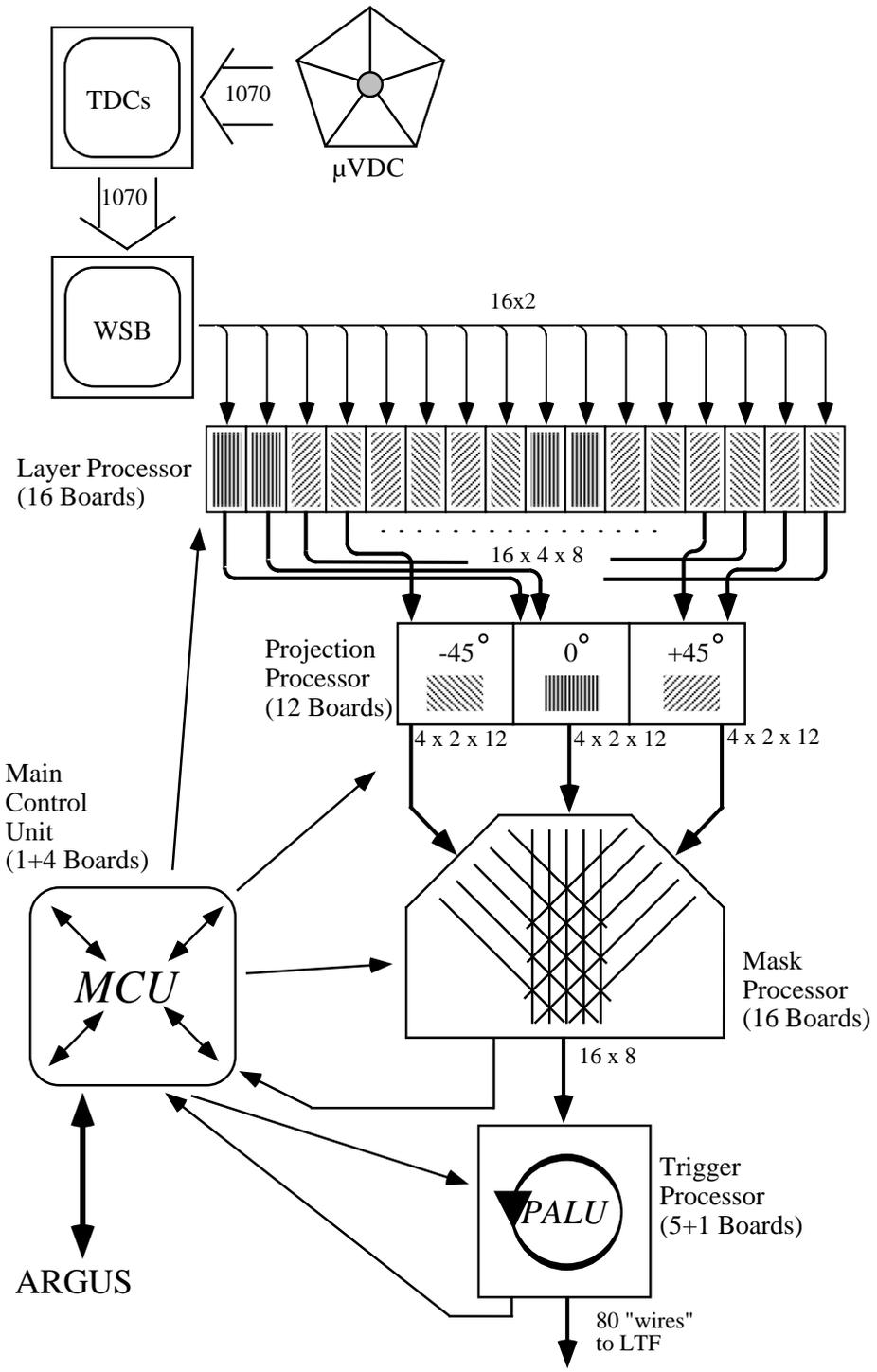

**Figure 6**   Schematic overview of the vertex trigger.

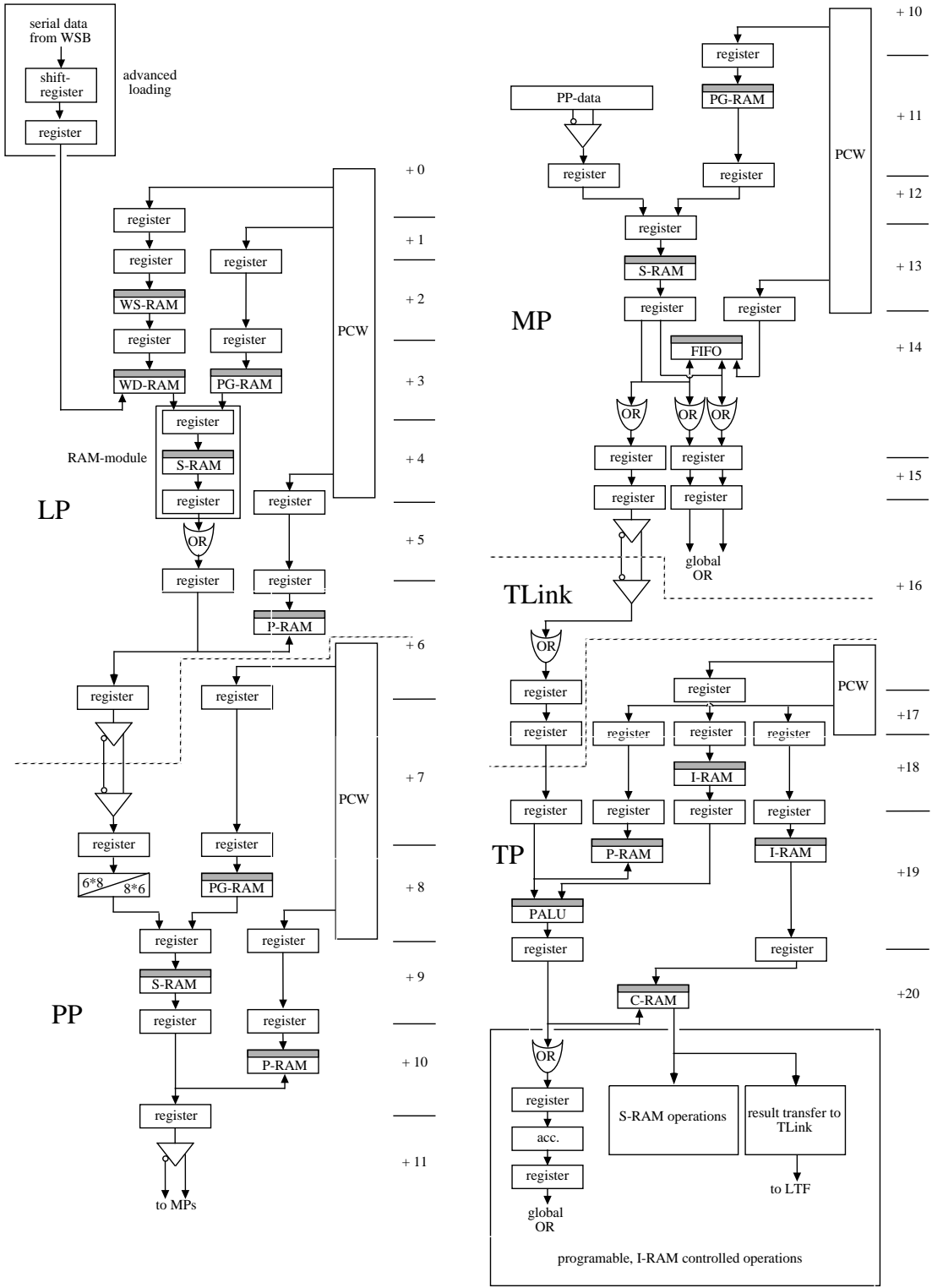

**Figure 7**   Data flow in the pipeline of the vertex trigger.

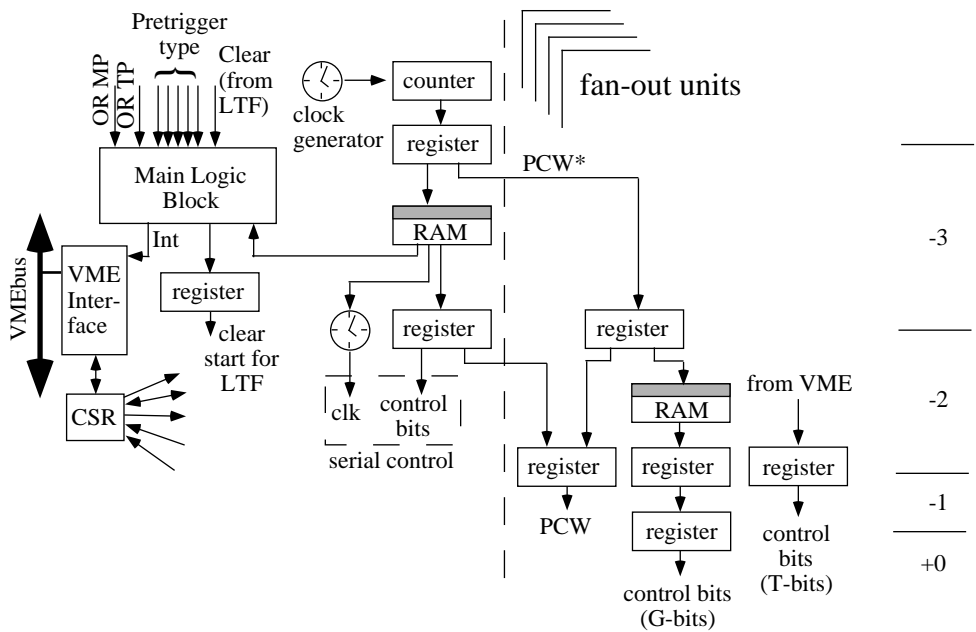

**Figure 8**  Block diagram and data flow of the Main Control Unit (MCU).

### 4.2.1 The Main Control Unit

The main control unit (MCU, figure 8) is responsible for the proper synchronisation of the pipeline and of the communication to and from the central trigger electronics of the experiment and the VME CPU. The MCU consists of two clock oscillators with programmable frequency, a control and status register (CSR), a finite-state-machine (Main Logic Block) to control the pipeline processing steps, the central program counter with trigger dependent ranges and fan-out units for every processor type to distribute the clocks, a pipeline (or program) control word (PCW, 13 bit), status bits (2), test mode bits (3 T-bits) and dynamic gate bits (2 G-bits).

Every four processor modules are connected via one flat ribbon cable on the VME J2/P2 connector to a front panel connector of the corresponding fan-out unit of the MCU. These fan-out units are again tightly connected via J2/P2 to the master board of the MCU that contains most of the logic and the VME interface.

Initialization of the trigger is done in the VME access mode. During run mode with inactive pipeline only the CSR may be written and some data banks read. After receiving a pretrigger or a command issued via VME the program counter is started with a start address depending on the pretrigger type. All circuitries with a power-saving option are powered up and the serial shift of the wire data is started. During this time the dynamic gate bits are used to tell the Layer Processor to store the arriving data in the wire data RAMs at the rate of the shift clock. Some cycles with null operations are used to clear all registers in the chain and the MCU before the first data are processed. The MCU accepts during the cycles three types of response from the trigger:

- A global OR from the Mask Processor during mask processing time

- Two types of global OR from the Trigger Processor at different times

Again depending on pretrigger type and programming these responses may trigger a start signal to the LTF to proceed and then issue an interrupt to the VME CPU to start the read-out procedure. A negative result will invoke an overall clear to all detector components. After completion all components are brought back into power-saving mode and the resulting data are available for read-out.

All time-dependent control signals of the MCU are programmed in RAMs addressed by the PCW. In addition all processor specific timings may be adjusted within a small margin of $\pm 8$ ns to correct for different cable lengths. The static test mode bits (T-bits) may be set for each processor type separately to specify different data sources and data storage modes. The protocol RAM (P-RAM) on the boards may then act as data input to complete the built-in logic analyser features of the system.

The clock frequency generated by the MCU pipeline clock was chosen to be 25 MHz.

### 4.2.2 Serial Loading

The signals of the $\mu$VDC drift cells are latched for each pretrigger into the TDC [9] 'hit registers'. The data from the 1070 TDC hit registers are sorted for each layer on the wire sorting board (WSB), which is a hardwired network to rearrange them in geometrically sequential order. The ordered sequence is then transmitted to the Layer Processor located in 20 m distance. To keep the number of cables small, the data are transmitted serially to the layer processor modules using 16 parallel/serial converters. Each converter sends the signals of one layer via two BNC cables to one processor module. To provide a continuous pattern everywhere in $\phi$, after a full cycle in one layer, the first 24 wires are transmitted again. Since the number of wires per layer varies between 35 and 100 (see section 2.2) the number of bits per layer varies between 59 and 124. This transmission scheme allows to store the complete set of all regions of 24 neighbouring wires in the Layer Processor.

Using a 30 MHz clock the serial loading is not synchronized with the pipeline. For a single event the whole data transfer takes about 4.7 $\mu$s.

### 4.2.3 Layer Processor

The Layer Processor is divided into 16 modules, each dedicated to one chamber layer. For the whole mask field processed in a cycle the modules determine one bit for each wire group from the logical OR of the wires belonging to that group. In one cycle this OR can be formed for 32 groups which is sufficient to cover the field of 512 masks.

In the input section the received serial data is transformed back to parallel data words: all connected regions of 24 neighbouring wires, each one shifted by one wire, are stored into the wire data (WD) RAMs such that they can be addressed during the different pipeline cycles.

At the start of the pipeline the cycle number selects in the wire section (WS) RAM the address of a region of 24 wires which covers the mask field processed in the given cycle. The 24 bit from the wires form together with a page number (5 bit selected in the PG-RAM) the address for the S-RAM. The output of the S-RAM is a 32 bit word which contains for 32 groups the information which of the groups are validated or not. With the 5 bit of the page number it is possible to define 32 different group configurations allowing, e. g., to take wire inefficiencies into account.

The results are then transferred to one of the three Projection Processors and at the same time to a protocol memory (P-RAM in fig. 7).

### 4.2.4 Projection Processor

The input to the Projection Processor are the 16 words, each 32 bit long, from the Layer Processor. The words, containing the wire group information are processed for each projection separately. The output is the projection quality word (3 bit) defined in table 1 for each projection and for each mask in the mask field.

The 32 bit words of the four $0°$-layers and of the six $\pm 45°$-layers are rearranged at the beginning of the process to 4 bit words ($0°$) and 6 bit words ($\pm 45°$), respectively. This information is reduced from 4 or 6 bit to a 3 bit quality number. The quality numbers are stored in RAMs (one RAM for each group) which are addressed by the 4 or 6 bit of a group and 8 bit for a page selected via the PCW (cycle number).

The hardware of the three Projection Processors is divided into four modules each leading to a total of 12 Projection Processor modules. Each module processes eight groups, i. e. $8 \times 4$ bit or $8 \times 6$ bit, respectively. This scheme allows parallel processing of the 32 possible groups of a mask field in all three projections. The output for each module are $8 \times 3 = 24$ bit which are arranged in two blocks of 12 bit words leading to eight words with 12 bit per projection. These $3 \times 8 \times 12 = 288$ bit are send to the Mask Processor.

### 4.2.5 Mask Processor

The Mask Processor combines the 3 bit projection quality numbers from the three projections to determine the 2 bit track quality numbers as defined in table 2.

Figure 9 shows the 512 masks of a mask field. Each mask is represented by a RAM which is addressed by the 3 bit quality numbers of each projection. In addition to these 9 bit the address contains 3 bit to select one out of 8 possible pages.

The field of 512 masks is divided into 16 sub-fields, each containing 32 masks, as shown in the middle part of fig. 5. The RAMs of each sub-field are installed on one processor module yielding a total of 16 Mask Processor modules [18]. If one mask is matched, the 2 bit quality numbers for each of the 32 masks is stored on a FIFO which can be read out when a trigger has stopped the experiment. A global OR of the two quality bits is formed separately. An open collector bus connecting all MP modules completes an OR over all masks in a cycle. These hits are sent to the MCU which accumulates during all cycles.

Further processing in the TP and LTF does not use $t$ information. Therefore a projection is performed in subsequent steps. The first step is accomplished on the MP modules. Each one transmits 8 bits to the TLINKboard.

### 4.2.6 TLink Board

The TLink board has a dual purpose:

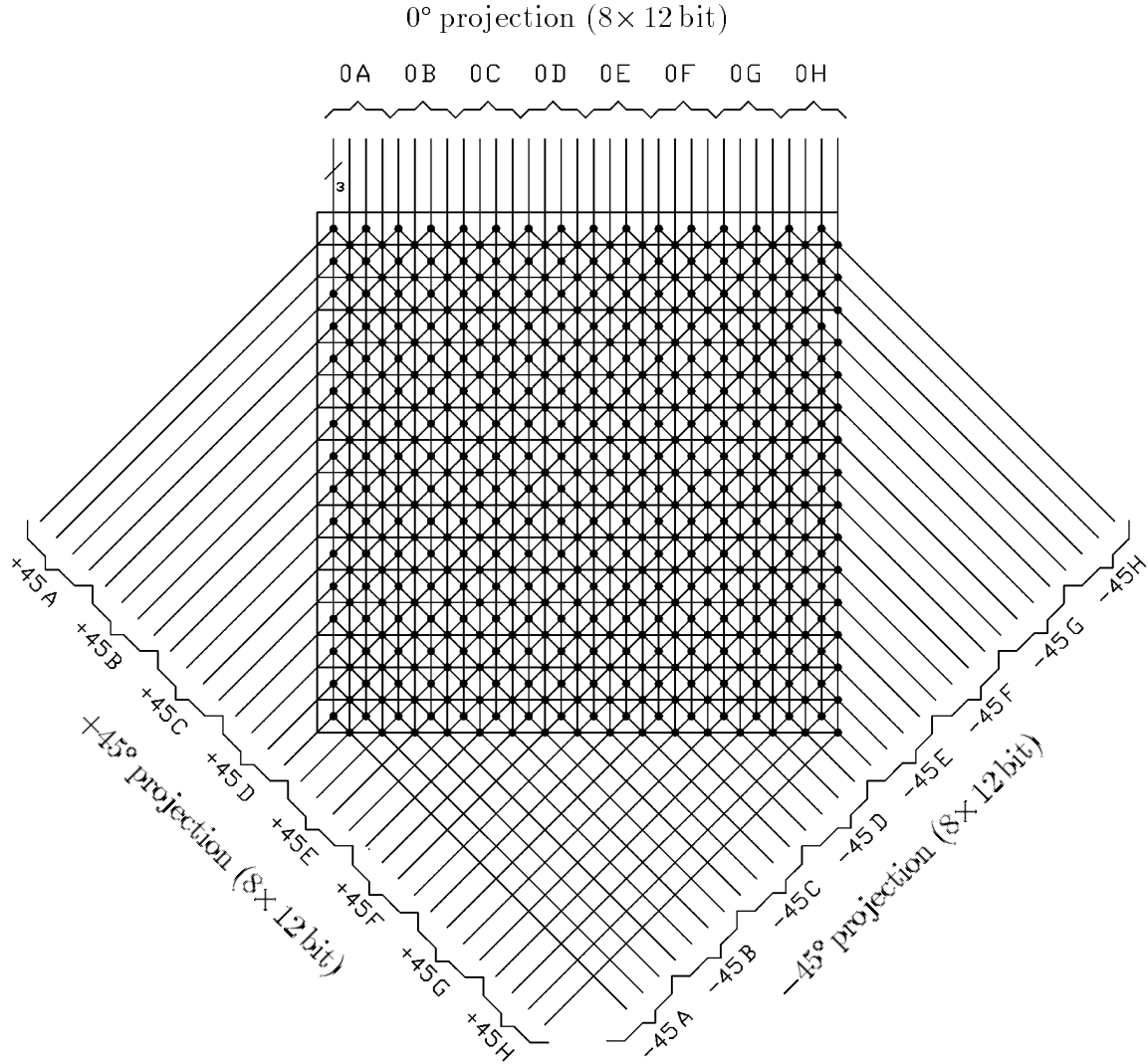

**Figure 9** Scheme of 512 mask RAMs (filled dots) in a field, built of two shifted 16×16 fields; Each RAM has individual 9 bit address lines, 3 bit from each projection (perpendicular and diagonal lines).

First, it receives the 16 × 8 bits of the projected quality 1 bits and forms an OR of groups of four corresponding to the same $s$ coordinates. The resulting 32 bits are presented on a common bus to all Trigger Processor boards.

Second, after completion, the Trigger Processor feeds the LTF reference hit pattern back to the TLink board as 5 chunks of 16 bits, which are filled into 80 hit registers for use of the LTF.

### 4.2.7 Trigger Processor

The Trigger Processor determines hits in the virtual reference layer of the LTF from the 0°-projections of the found tracks, and searches for pairs of tracks with at least 90° separation in azimuth $\phi$ coming from the same vertex region.

These complex tasks are performed on five 'special purpose computer' boards, each executing a program in phase with the pipeline (see fig. 7). It is capable to use mapping functions stored in look-up tables (S-RAM), and in parallel do logical operations with data on the two data buses (A and B) in a primitive logical unit built with a PAL (PALU), which has a set of 7 different instructions. Intermediate results are stored in a 4kB memory (C-RAM). The processing sequence is stored in an instruction RAM (I-RAM). Since the processing steps require several clock cycles, the task is shared between five identical processor boards. For the final steps, these boards can communicate via the I/O bus shared with the TLink feed. A second output line is used to send a trigger signal to the MCU as an OR of the PALU output register bits. Details can be found in [10].

The whole program is written in an assembler language [11], and is translated into an I-RAM sequence by a cross-assembler on the host computer. The instruction set is listed in table 3.

Functions are realized as mappings via the S-RAM. Each of the four bytes of a 32-bit word form part of an address, which points to a 32-bit wide S-RAM location. Another 5 bit identify the map (function), and are part of the instruction. Finally, the four data words at the corresponding locations are ORed, and available on the A data bus on request. All function maps are calculated during the trigger initialization on the host computer, and transferred into the S-RAM of the Trigger Processor boards.

In the first step of the processing sequence, all masks with the same $s$ coordinate and within 3 neighbouring groups in $z_0$ are ORed, completing the projection started on the TLink board for tracks occurring in parallel.

Input to the five Trigger Processor boards is provided sequentially, starting at sector 0, in the following order (4 sequential cycles give results on four fields lined up in $t$ direction, which are always ORed):

```
cycle:         0  4  8   12 16 20 24 28 32 36 40 44   48 52 56 60 64 68 72 76 80 84 88 92   96 100 104...
sector/field:  0/2,3,4,5                              0/A,B,C,D                             1/2,3,4,5...
z0 group:      0  1  2  3  4  5  6  7  8  9  A  B    0  1  2  3  4  5  6  7  8  9  A  B    0  1  2  3  4  ...
T 0            ××× ×××       ××× ×××       ××× ...
T 1              ××× ×××       ××× ×××       ××× ...
T 2                ××× ×××       ××× ×××       ××× ...
T 3                  ××× ×××       ××× ×××       ×× ...
T 4                    ××× ×××       ××× ×××       × ...
               low s half           high s half           low s half
```

A cross (×) indicates, that the corresponding Trigger Processor accepts input during the four cycles corresponding to four mask fields.

After the OR of all aligned mask patterns during the input phase, clusters of neighbouring masks in the $s$ projection are combined and mapped via their central $s$ coordinate onto LTF reference cells. The reference layer has 80 cells of equal size in $\phi$, 8 cells per half sector. The mapping of the 32 masks to 8 bits in one half sector is determined by the geometrical requirements: a cell is set if the centre of gravity of a cluster of masks is within the cell boundaries, and clusters with 10 or more masks in $s$ direction set two neighbouring cells. For this calculation, the mask pattern is shifted (via three special mappings) to have a cell boundary at the centre of a region of eight masks, stored in one byte. Six regions are formed in each half sector, three more are formed at the centre of a sector from bits of both halves. Bit 0 and 7 of a region's byte are a logical AND of the corresponding mask bit with the next one outside the region, except the two mask bits next to the vane, which are ORed to form bit 0 of the first region. After this geometrical rearrangement, there is a unique mapping of the 256 possible values of the byte of each region to the two bits of the associated two reference layer cells defined by the geometrical requirement given above.

**Table 3**  Trigger Processor instruction set. The PALU input is from two buses A and B, its output register is R. c denotes a C-RAM address (10 bit for 1024 32-bit words), s are five bits of an S-RAM address (selecting one of 32 functions) and simultaneously an output marker. Y is a register being written from the B bus. The source A has to be replaced by I (input at I/O port), Y (B bus register Y), S (function value from S-RAM) or P (protocol P-RAM, for test purposes). Instructions from the six different groups can be executed in parallel, unless there is an address conflict.

| instruction | clock cycle | action |
|---|---|---|
| Y=B[c] | 1 | CRAM→B→Y |
|  | 2 | (Y→A possible) |
|  |  | *PALU operations:* |
| R=R | 1 | R→R |
| R=A | 1 | A→R |
| R=B[c] | 1 | CRAM→B, B→R |
| R=-B[c] | 1 | CRAM→B, ¬B→R |
| R\|=A | 1 | A∨R→R |
| R=A\|B[c] | 1 | CRAM→B, A∨B→R |
| R=A&B[c] | 1 | CRAM→B, A∧B→R |
|  |  | *PALU R-bits global OR (accumulated):* |
| OR | 1 | define R |
|  | 2 | OR step 2 |
|  | 3 | accumulate |
| GOR=0 | 1 | clear global OR |
|  |  | *C-RAM write:* |
| B[c]=R | 1 | R→CRAM→B |
|  |  | *S-RAM function:* |
| S=f[s](B[c]) | 1 | CRAM→S-RAM address |
|  | 2 | S-RAM step 2 |
|  | 3 | S-RAM step 3 (OR 4 words) |
|  | 4 | (S-RAM→A possible) |
|  |  | *pattern output:* |
| OUT(s) | 1 | A→I/O port |
|  | 2 | provide s = sector number |
|  | 3 | validate s via bit change |

The azimuthal two-track correlation for the second level trigger decision requires functions, which provide bit masks for all possible hits with a minimum separation of $\Delta\phi > 90°$ from the hit cells, calculated in the LTF cell resolution for all groups of three neighbouring $z_0$ patterns. These functions are computed assuming a cyclic string of 80 units, by setting all bits more distant than $n_{\min} = 20$ cells from a hit cell. An AND of these bit-masks with the found track pattern leaves no bit set, if and only if there is no pair of tracks with $\Delta\phi > 90°$.

The programs in all five processor boards are almost identical, with two exceptions:

(i) The start of the program is shifted by four clock cycles from one processor to the next in order.

(ii) The final combination of all results is performed in processor T0, which forwards the LTF reference cell information and provides the single track and coincidence trigger signals. Depending on the set pretriggers, these signals are acknowledged or ignored by the MCU.

The trigger pipeline starts 149 cycles after activating the pipeline clock, i.e. with a delay of about $6\,\mu$s, to allow for a safe serial feeding from the wire-sorting board. The first data arrive at the entry register of the Trigger Processor 17 cycles later. The first trigger signal is sent out at cycle 671, the two-track coincidence signal at cycle 695. Thus, the trigger decision is available $28\,\mu$s after the start.

## 5 Trigger Control

To aid initialization, testing and read-out all VT Processors are housed in VME bus crates. The processor boards are equipped with A24/D32 VME interfaces. The MCU master VME interface is further capable of requesting VME interrupts and providing the required interrupt vector/status information.

The VME system architecture as sketched in fig. 5 uses the CES VMV bus as VME intercrate connection. Each VME crate is connected to the VMV bus via a VIC8250 [12]. The whole system is served by an Apple Macintosh host computer with a Motorola 68020 CPU which is connected via its NuBus [13] and a MAC7217 [14] to the VMV bus. The interface to the ARGUS online system is realized as a mailbox between the ARGUS CAMAC system and the VMV bus. The information needed from the first level trigger is accessible via the VME housed CAMAC branch driver CBD8210 [15]. To manage online data transfer a FIC8230 [16] VME CPU is incorporated into the system. It is a Motorola 68020 based single slot crate computer with 2MB dual ported static RAM accessible from the CPU and from VME.

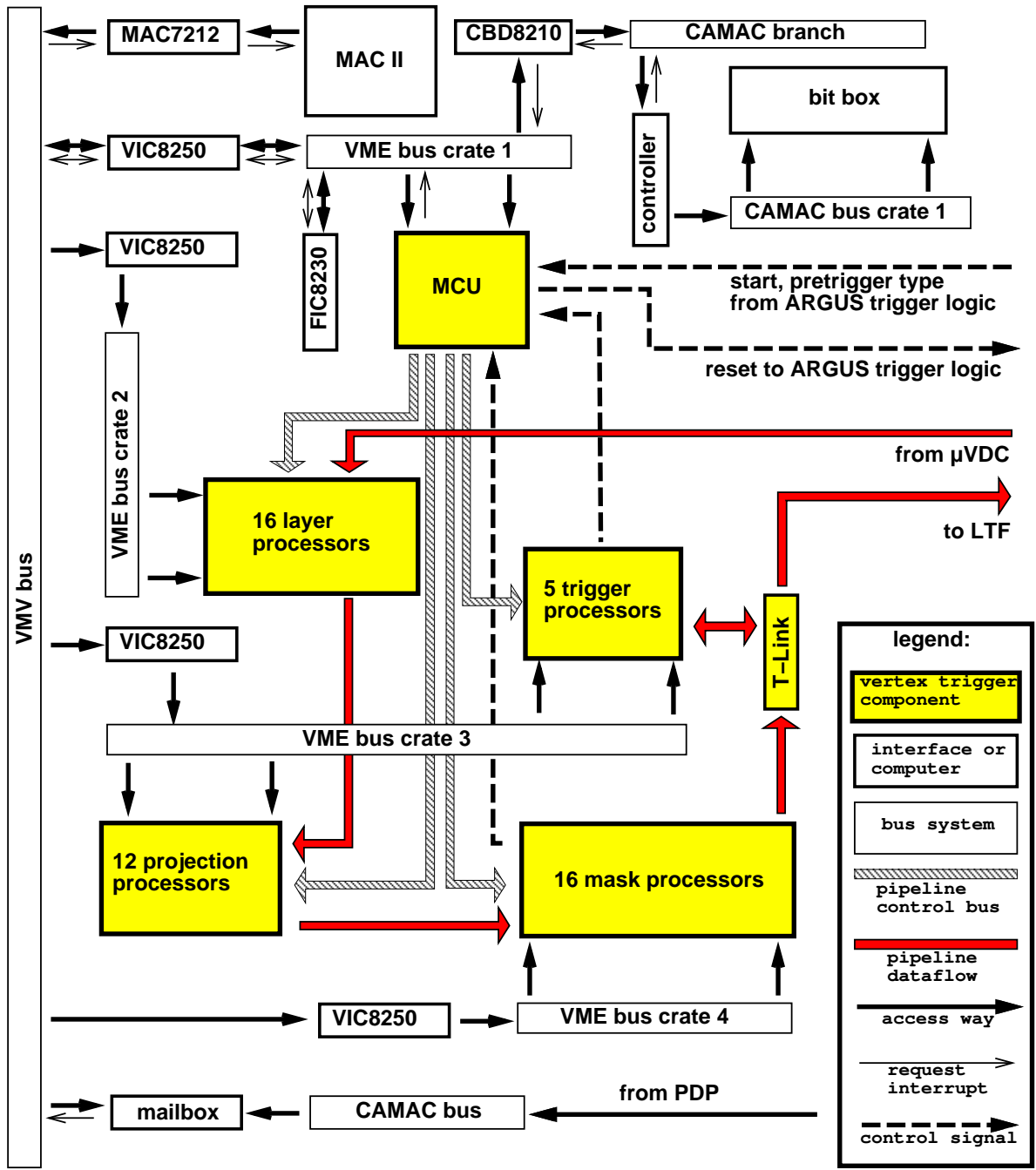

**Figure 10** Block diagram of the trigger control system

The initialization phase starts with the booting of the host computer. From the host all interfaces are set up, the VME CPU is booted and an executable program is downloaded. The interfaces are configured to map logical addresses on the host to the same physical storage addresses as addresses on the VME CPU with the exception of
the global accessible RAM on the VME CPU. All the VT look-up tables realized as
static random access memory are loaded upon initialization. Finally the whole system is
enabled by writing the required control bits into the MCU master unit.

Tests of the system are possible by the built in 'logic analyser':

- In each cycle the bit pattern at each processor's output is stored in protocol RAM (P-RAM).

- The P-RAM may act as a data source for the following pipeline steps.

- The wire pattern of the $\mu$VDC may be written into the storage of the Layer Processors directly.

- The pipeline can be configured to run in any specified pipeline address range.

Using these features process debugging/testing on the step by step level is possible. The
sole exception is the wire transfer from the TDC hit registers to the Layer Processors: The
TDCs allow only to set all even or all odd hit registers before starting the transmission.
Therefore only a transmission test but not a wire pattern dependent test is possible.

## 5.1 Control Software

Three major software tasks can be identified: Initialization of the whole system, monitoring and test of the operation and control of the running and read-out of trigger data. The initialization and test routines run on the host computer. For the other tasks the host serves as user interface while the execution takes place on the crate computer whose programs are down-loaded by the host as well. All software has been developed on the host computer using the 'Macintosh Programmers Workshop' (MPW) development system.

The host computer runs under a windows based operating system. This comfortable user interface is used extensively. An application was developed which includes all code needed and provides terminal I/O and histogramming features. All VT parameters are set via menues in this program. External (VME) interrupts are served and a small set of interrupt driven communication commands between the host computer and the crate computer is defined: transfer of data, printout from the crate computer, and terminal input for the crate computer.

The crate computer runs with a rudimentary operating system: The small kernel allows starting of programs at an externally specified address. It also provides memory management and terminal I/O routines. A tool running on the host computer was developed which allows the loading and starting of programs compiled on an Apple Macintosh. The VME system setup allows to run routines developed and tested on the host computer to run without change on the crate computer.

### 5.1.1 Initialization and Test

The parameters of the VT processor boards are set by initializing look-up tables. The used storage amounts to 32 Mbit. The parameters of the trigger algorithm are few:

the width of a mask in a $\mu$VDC layer, the number of vertex positions covered, the $\cot\theta$ boundaries of the active volume, the number of required hits along a track path dependent on its position in the track parameter space and a dead wire map. The trigger algorithm parameters are mapped into the hardware parameters by the initializing software.

Tests are performed after the initialization on all hardware units using the protocol RAMs: A wire pattern is loaded into the Layer Processors, the pipeline is triggered by a write access to the MCU and the protocol RAMs are read out. The protocols are then compared with the simulated results to detect possible malfunctions.

### 5.1.2 Read-Out and Run Control

The pretrigger rates of up to several 100 Hz and the requirement of fast reenabling of the system after a CLEAR calls for a very fast response on interrupt requests to the crate computer. This is implemented by running the processor with a dedicated operating system and using the multilevel hardware interrupt (IRQ) scheme of the Motorola 68020 processor.

On successful completion of trigger processing the MCU interrupts the online processor. The CPU reads the mask data via VME and VMV from the FIFOs of the Mask Processors and the cell pattern from the C-RAM of the Trigger Processor T0 into its memory and transfers them to the CAMAC mailbox where they are taken over by the ARGUS online system.

A CLEAR condition may occur at any time from other components in the online system. Each CLEAR is transmitted via an IRQ of the highest possible priority (7) from the MCU to the crate CPU which suspends all read-out processing by resetting the corresponding stack pointers: Program execution continues in the normal program independent of the depth of nested interrupt requests. A read-out request uses an IRQ of level 6 and may hence be interrupted (and aborted) by a CLEAR condition.

Error detection is provided by monitoring mask occurrences which are read from the Mask Processors both at VT acceptance or rejection. The latter is possible via the VETO trigger or the ETOT or random pretriggers. Further information such as $\mu$VDC wire maps on the Layer Processor inputs is held on the online processor and is available on user request from the host computer.

Two buffers of event data are available on the crate computer. The first is used to transmit data to the ARGUS online system, the second to supply monitoring information. If the monitor task has finished with a buffer the two buffers are swapped: the monitoring uses the data currently available in the read-out buffer, while the read-out fills the other buffer with subsequent events. This operation allows for rather independent and asynchronous operation of both tasks. The stack resets of the read-out tasks do not affect the monitoring since both tasks are using different stacks and stack pointers.

# 6 Performance of the Vertex Trigger

The Vertex Trigger was used for the first time in the 1991 running of the ARGUS experiment. The hard- and software was working properly and allowed a stable operation. The pipeline could be run with the design value of 25 MHz so that the total decision time of the VT was about 28 $\mu$s.

**Table 4**  Single track efficiencies determined for Bhabha events.

| track quality | whole chamber | plateau of chamber |
|---|---|---|
| 3 | 0.81 | 0.97 |
| 1 | 0.85 | 0.98 |

An initial test run where the VT results were recorded but the event read-out was done independently of the VT allowed to make an unbiased determination of the VT efficiencies and rejection power. The recorded trigger information of the VT was compared to the corresponding results for each event from a simulation program using the available $\mu$VDC data. The comparison showed that the trigger hard- and software is well understood [17, 18].

Using events triggered by ETOT and VETO pretriggers one can obtain data where the taken VT and LTF decisions are recorded but not used. Using such events of the whole run the track efficiencies have been determined for track qualities 3 and 1 (see table 2). The results obtained for Bhabha events are listed in table 4. Bhabha events were chosen because of their clean signature, but the results hold for any $e^+e^-$ annihilation event. Since there are inefficiencies due to the chamber geometry in the region of the vanes the track efficiencies are given both as the average over the whole chamber as well as for the plateau region between the vanes. However, even in the plateau the inefficiencies in the $\mu$VDC affect the track finding since its mean plateau efficiency is only 96%. All trigger inefficiencies found are for tracks with missing hits in several layers of the chamber. The masks of the Vertex Trigger cover the complete parameter volume of the interesting tracks.

Also the dependence of the single track efficiencies on the track parameters was found to agree with the simulations. Full efficiency was reached for $p_T = 60$ MeV, the lowest momentum which can be reconstructed in the main drift chamber. For the impact parameter $d_0$ full efficiency was found for $d_0 < 3.7$ mm falling off to 50% at $d_0 < 7.3$ mm. The 12 $z_0$ mask sets covered with full efficiency the range $-42$ mm $< z_0 < +42$ mm. The track parameter resolution (full width at half maximum) of each mask agreed with the expectation: $\Delta \phi = 20$ mrad, $\Delta \cot \theta = 0.3$, $\Delta d_0 = 7.3$ mm and $\Delta z_0 = 14$ mm.

The trigger efficiency and background rejection has been determined for two trigger conditions:
 condition 1:   1 track found in VT
 condition 2:   2 tracks found in VT with opening $\Delta \phi > 90°$

At a pretrigger rate of 120 Hz the background rejection with the combined LTF-VT trigger was 10 and 15 for condition 1 and 2, respectively. The corresponding reduction factors due to the VT alone were calculated to be 2.6 and 3.5. The efficiencies for two-prong events were 99% and 76% and for multi-hadronic events 100% and 97% for condition 1 and 2, respectively. In order to obtain large efficiencies only condition 1 was used during data taking. This resulted in a deadtime of the experiment well below 25% which is an improvement by a factor 2. The performance in terms of 'Vertex Finding' is best illustrated by figure 11. Only events with tracks from the main vertex region have a large detection efficiency. Note that all events in this plot have passed the old trigger setup including LTF. The long tails of the hatched distribution (with VT) are mostly due to background events with a number of hits in the $\mu$VDC too large to discriminate tracks from accidental combinations.

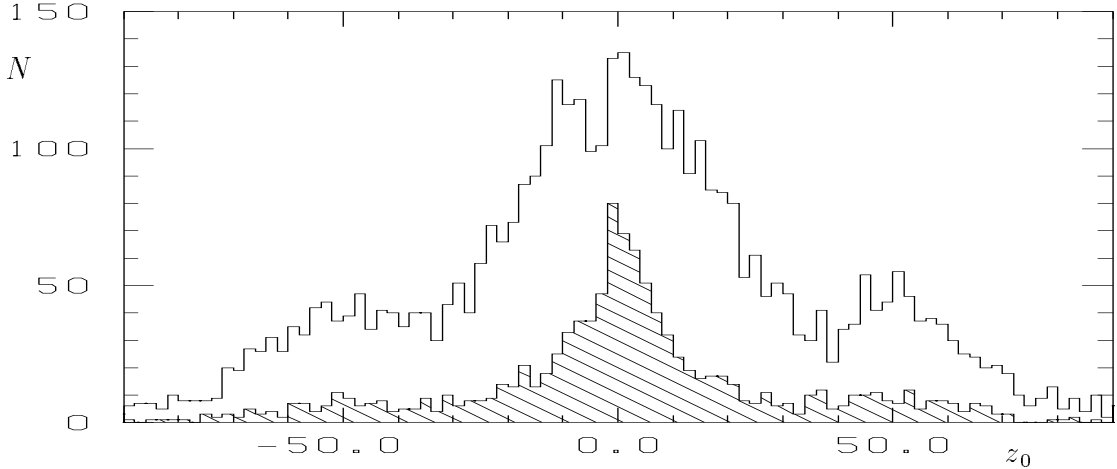

**Figure 11** Distribution of $z_0$ of the track with minimal $|z_0|$ of all tracks in an event without and (hatched) with Vertex Trigger.

The obtained track efficiencies were parametrized as a function of the 5 track parameters and used to calculate event efficiencies from the contributing tracks. This calculation agreed well with the directly measured efficiencies for multi-hadronic and Bhabha events, indicating that the measured single track efficiencies are not dependent on the event type, and not dependent on the event multiplicity. This method allows a high precision trigger efficiency determination even for low statistic data samples since the single-track efficiencies are obtained from other data.

With this initial experience improvements in the mask definitions have been worked out which allow to increase the VT rejection factors to 4.5 (using condition 2), and at the same time to even gain somewhat in the efficiency [17].

## 7  Summary

A fast vertex trigger processor has been developed for the ARGUS experiment to reduce the trigger rate caused by tracks not originating from the interaction region. This trigger improvement became necessary after the installation of a new vertex detector with a beam pipe of only 10.5 mm radius. The Trigger Processor uses the wire hit information of the ARGUS Micro Vertex Drift Chamber ($\mu$VDC) to recognize tracks originating from the interaction region within $d_0 \leq 7$ mm radially and $\Delta z = \pm 42$ mm axially.

Using the 16 wire layers of the $\mu$VDC with $0°$, $+45°$, and $-45°$ stereo wires allows a full three-dimensional track recognition. The processor compares the hit wires with 245760 masks defining tracks from the interaction region. The masks are processed in 480 serial cycles of a synchronous pipeline. Each step deals with 512 masks in parallel. Employing a 25 MHz clock the total processing time is about 28 $\mu$s.

The trigger reduces the background rate by a factor 2.5 in the first setup, and can be tuned to a reduction factor 4.5. The single track trigger efficiency for Bhabha events was determined to be 97% if the track passes through a maximally efficient part of the $\mu$VDC.